\documentclass[sigconf]{acmart}

\usepackage{amsmath,amssymb,amsfonts}
\usepackage{graphicx}
\usepackage{textcomp}
\usepackage{xcolor}

\usepackage{multirow}
\usepackage{algorithm,algorithmicx,algpseudocode}
\usepackage{hyperref}
\usepackage{comment}
\usepackage{array}
\usepackage{pifont}
\AtBeginDocument{
    \hypersetup{
        colorlinks=true,
        citecolor=ACMPurple,
        urlcolor=black,
        linkcolor=ACMPurple
    }
}
\usepackage{setspace} 

\copyrightyear{2026}
\acmYear{2026}
\setcopyright{cc}
\setcctype{by}
\acmConference[ICCAD '26]{IEEE/ACM International Conference on Computer-Aided Design}{November 08--12, 2026}{San Jose, CA, USA}
\acmBooktitle{IEEE/ACM International Conference on Computer-Aided Design (ICCAD '26), November 08--12, 2026, San Jose, CA, USA}
\acmDOI{10.1145/3831252.3834075}
\acmISBN{979-8-4007-2873-0/2026/11}

\begin{document}

\ccsdesc[500]{Hardware~Electronic design automation}

\title{APEX-RBD: Mixed-Precision Exploration Framework for Hardware-Efficient Robot Dynamics Accelerator Design}


\author{Xingyu Liu}
\email{xliugu@connect.ust.hk}
\orcid{0009-0006-0897-8670}
\affiliation{%
  \institution{The Hong Kong University of Science and Technology}
  \state{Kowloon}
  \country{Hong Kong}
}

\author{Hanwei Fan}
\email{hfanah@connect.ust.hk}
\orcid{0000-0002-1177-2108}
\affiliation{%
  \institution{The Hong Kong University of Science and Technology}
  \state{Kowloon}
  \country{Hong Kong}
}

\author{Chaofang Ma}
\email{cmaaw@connect.ust.hk}
\orcid{0009-0009-2082-4916}
\affiliation{%
  \institution{The Hong Kong University of Science and Technology}
  \state{Kowloon}
  \country{Hong Kong}
}

\author{Jiawei Liang}
\email{jliangbr@connect.ust.hk}
\orcid{0000-0003-0748-3183}
\affiliation{%
  \institution{The Hong Kong University of Science and Technology}
  \state{Kowloon}
  \country{Hong Kong}
}

\author{Guangyu Hu}
\email{ghuae@connect.ust.hk}
\orcid{0000-0001-5077-8361}
\affiliation{%
  \institution{The Hong Kong University of Science and Technology}
  \state{Kowloon}
  \country{Hong Kong}
}

\author{Jiang Xu}
\email{jiang.xu@hkust-gz.edu.cn}
\orcid{0000-0001-9089-7752}
\affiliation{%
  \institution{The Hong Kong University of Science and Technology (Guangzhou)}
  \city{Guangzhou}
  \country{China}
}

\author{Wei Zhang}
\authornote{Corresponding Author.}
\email{wei.zhang@ust.hk}
\orcid{0000-0002-7622-6714}
\affiliation{%
  \institution{The Hong Kong University of Science and Technology}
  \state{Kowloon}
  \country{Hong Kong}
}

\begin{abstract}

Rigid Body Dynamics (RBD) forms the computational core of real-time robotic control, but its immense computational complexity creates a performance bottleneck that necessitates dedicated hardware accelerators. However, the substantial hardware resource and power costs of these accelerators make their deployment on resource-constrained edge platforms highly challenging. While quantization offers a promising path to optimize RBD hardware for edge computing, existing uniform-precision approaches remain inefficient by ignoring the diverse quantization sensitivities of different variables. Although mixed-precision offers a superior alternative, its exploration is intractable due to a vast search space and the prohibitive cost of closed-loop simulation for motion accuracy evaluation.

To address these challenges, we introduce APEX-RBD, an automated framework that makes mixed-precision exploration computationally tractable while effectively identifying hardware-efficient configurations. Specifically, it performs physics-driven search space pruning via variable grouping and sensitivity analysis, and employs a data-efficient, prior-informed surrogate model to enable rapid trajectory error prediction. This formulation guides a hybrid optimizer to identify area- and power-efficient designs under user-defined accuracy and performance constraints. Experimental results demonstrate that APEX-RBD discovers designs achieving up to \textbf{1.9$\times$ area reduction} and \textbf{1.8$\times$ power savings} compared to uniform-precision baselines across diverse robotic platforms.

\end{abstract}

\vspace{-3pt}
\keywords{Robot Rigid Body Dynamics, Mixed-Precision Quantization}

\maketitle
\section{Introduction} \label{section_1}

The demand for autonomous robots necessitates kHz-level real-time control on edge devices~\cite{Robomorphic,liuDRACOCodesignDSPEfficient2025a}. Central to this control loop are Rigid Body Dynamics (RBD) computations~\cite{RBDAlgorithms}, which must be executed at every time step. However, their high computational cost makes them a critical performance bottleneck, as they are dominated by serial multiply-accumulate (MAC) operations and consume up to 90\% of a controller's runtime~\cite{AnalyticalDerivativesRSS,plancherPerformanceAnalysisParallel2020}. While dedicated RBD accelerators are limited by finite on-chip resources~\cite{DaduRBD,Roboshape}, quantization offers a promising path to boost performance~\cite{liuDRACOCodesignDSPEfficient2025a}. It allows shrinking the physical size of MAC units, enabling more of them to operate in parallel within the same hardware footprint, thereby increasing throughput. 

While quantization has achieved widespread success in optimizing deep neural networks (DNNs) for robotic perception~\cite{parkSaliencyAwareQuantizedImitation2025,paniegoModelOptimizationDeep2024,zhang2026versaq3darchitecturesupportvisual}, applying this technique to RBD within the control loop introduces different and more severe challenges. Unlike perception-oriented DNNs, which possess inherent statistical robustness to numerical noise and typically process open-loop data frame-by-frame, RBD computations dictate physical actions within a continuous feedback loop~\cite{RBDAlgorithms}. A minor quantization inaccuracy in calculating a joint torque physically alters the robot's actual trajectory. This creates an erroneous state that is fed back into the controller's highly non-linear equations (e.g., recursive trigonometric chains and matrix inversions~\cite{carpentierAnalyticalInverseJoint}). Consequently, these errors iteratively compound and amplify over time, leading to trajectory drift and system instability~\cite{liuDRACOCodesignDSPEfficient2025a}. This complex, system-level error propagation presents a major barrier to applying quantization in robotic control.

The recent work, DRACO~\cite{liuDRACOCodesignDSPEfficient2025a}, represents a pioneering effort in RBD quantization. However, its quantization framework is specifically designed for FPGA deployment. Although FPGAs can support heterogeneous bit-width arithmetic, MAC-dominated RBD accelerators rely heavily on pre-fabricated Digital Signal Processing (DSP) slices~\cite{DSP48,FLEX,FPGN} for performance and energy efficiency. Because these DSP slices have fixed native multiplier dimensions, arbitrary mixed-precision assignments may underutilize DSP resources or spill into LUTs, making hardware savings non-proportional to reduced bit-widths. Therefore, DRACO adopts a uniform-precision approach, where all variables share the same bit-width. For the strict power and throughput requirements of real-time edge control, Application-Specific Integrated Circuits (ASICs) remain a compelling target platform~\cite{Robomorphic,wanRoboticComputingFPGAs2022}. At design time, ASICs can instantiate arithmetic units with customized bit-widths, enabling efficient specialization for a defined deployment workload envelope. Applying a uniform-precision approach on such an ASIC is inherently inefficient. This "one-size-fits-all" strategy squanders the ASIC's capability for bit-level customization, resulting in unnecessarily oversized data paths. Furthermore, it fails to exploit a crucial fact in RBD: a variable's sensitivity to motion accuracy loss and its hardware cost are independent factors. For instance, a frequently-used, low-sensitivity variable can be aggressively quantized for major hardware savings, while a sensitive but rarely-used one can retain high precision with minimal overhead.

\begin{figure}[t]
    \centering
    \includegraphics[width=\linewidth]{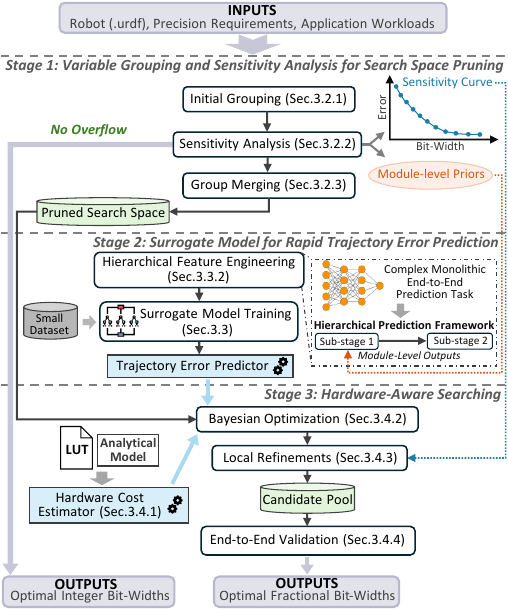}
    \vspace{-10pt}
    \caption{\textbf{Overview of the APEX-RBD framework.} It automates mixed-precision exploration across three stages. \textit{Stage 1} uses physics-driven grouping and sensitivity analysis to prune the search space and determine optimal integer bit-widths, preventing overflow. \textit{Stage 2} trains a hierarchical, data-efficient surrogate model to serve as a trajectory error predictor. Guided by this predictor and an analytical cost estimator, Stage 3 leverages Bayesian Optimization and local refinements to identify optimal fractional bit-widths.}
    \label{fig_quantization_framwork}
    \vspace{-10pt}
\end{figure}


This principle directly motivates \textbf{mixed-precision quantization}, a strategy that tailors the bit-width to each variable's characteristics. Importantly, our goal is not to identify a universal datatype set for all RBD scenarios; robot dynamics, workloads, and accuracy requirements vary across platforms. Instead, APEX-RBD searches for an efficient mixed-precision configuration within a user-defined deployment workload envelope, which may include multiple representative tasks or operating conditions. While this fine-grained approach avoids the significant resource waste of uniform quantization designs, it transforms the design problem into a vast combinatorial search. This leads to the core problem addressed in this work: \textit{how to navigate this massive search space to find the optimal mixed-precision configuration that minimizes hardware area and power consumption}. Crucially, this selected configuration must satisfy two constraints over the specified workload envelope: the robotic system's physical requirement for motion accuracy (i.e., maintaining trajectory fidelity without control instability) and the RBD accelerator's performance requirement for high throughput.


Finding the optimal configuration poses a significant challenge due to two factors. \textbf{First}, the search space is exponentially vast; for key RBD algorithms~\cite{RBDAlgorithms,plancherGRiDGPUAcceleratedRigid2022}, quantizing over 50 different variables with a range of 12 bit-widths yields a space exceeding $12^{50}$. \textbf{Second}, evaluating the motion accuracy of each configuration requires end-to-end closed-loop simulation~\cite{liuDRACOCodesignDSPEfficient2025a} to explicitly model the robot's physical dynamics and verify whether the accumulated trajectory error satisfies the required precision constraints. However, this process is prohibitively time-consuming. To illustrate, even with 8-thread parallelization, verifying 100 candidates takes 8.5 hours. This severe evaluation cost renders conventional search methods and data-hungry strategies like Reinforcement Learning~\cite{kwonRLPTQRLbasedMixed2024,ExplainableFuzzy} impractical for exploring the vast design space, as they often require thousands of simulation-based samples for convergence.

In this paper, we propose APEX-RBD, a framework to address these challenges. Our methodology tames this complexity in three stages. \textbf{First}, to prune the vast search space, APEX-RBD performs a two-step reduction, combining physics-based variable grouping with a sensitivity analysis to merge dimensions while preserving critical information. \textbf{Second}, to overcome the time-consuming evaluation bottleneck, we develop a surrogate model that predicts trajectory error in milliseconds, enabled by a novel hierarchical feature engineering approach. Instead of learning the complex end-to-end mapping directly, we decompose the problem into simpler, physically-grounded sub-problems and leverage the sensitivity priors gained from our first step. \textbf{Finally}, armed with this fast predictor and a hardware cost estimator, we employ a hybrid search strategy. It synergizes Bayesian Optimization for global exploration with hardware- and sensitivity-aware local heuristics for fine-tuning, efficiently identifying the near-optimal configuration under user-defined constraints.


Our contributions are summarized as follows:
\vspace{-3pt}
\begin{itemize}
    \item We propose the first mixed-precision quantization framework for RBD computations (Fig.~\ref{fig_quantization_framwork}) to find area- and power-efficient fixed-point configurations for deployment on ASICs, subject to user-defined motion accuracy and accelerator performance constraints.
    \item To make the vast search space tractable, we introduce a novel methodology that combines physics-driven variable grouping and sensitivity analysis for space pruning, with a few-shot, prior-informed surrogate model that accurately predicts trajectory error in milliseconds.
    \item To facilitate an efficient, hardware-aware search, we develop a cell-library-based analytical model for hardware cost estimation. This model guides a constraint-aware hybrid optimizer that combines Bayesian Optimization for global exploration with a heuristic local search for fine-tuning.
\end{itemize}
Experimental results demonstrate that APEX-RBD discovers designs with area and power reductions of up to 1.9$\times$ and 1.8$\times$ over uniform-precision baselines under identical accuracy constraints.

\section{Preliminaries} \label{section_2} 
\subsection{RBD Algorithms and Accelerator} \label{section_2_1}

\begin{figure*}[t]
    \centering
    \includegraphics[width=\linewidth]{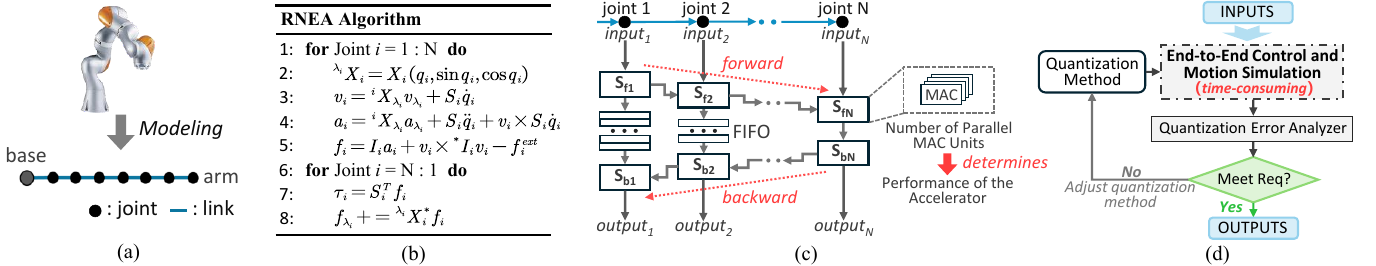}
    \vspace{-22pt}
    \caption{Overview of RBD concepts, RBD accelerator architecture, and quantization framework. (a) A robot arm modeled as a kinematic tree of links and joints. (b) The bidirectional traversal structure in the RNEA algorithm, consisting of a forward pass (joints 1 to N) and a backward pass (joints N to 1). (c) Accelerator architecture mapping the bidirectional dataflow, with performance determined by the number of parallel MAC units. (d) Uniform-precision quantization framework in DRACO.} 
    \label{fig_preliminary}
    \vspace{-5pt}
\end{figure*}

RBD computations are central to robotic control. Key algorithms, such as the Recursive Newton-Euler Algorithm (RNEA)~\cite{RBDAlgorithms} and Minv algorithm~\cite{carpentierAnalyticalInverseJoint,DaduRBD,liuDRACOCodesignDSPEfficient2025a}, operate on a model of the robot as a kinematic tree (Fig.~\ref{fig_preliminary}(a)). A notable computational pattern of these algorithms is the bidirectional data traversal (i.e., forward and backward passes), exemplified by RNEA's dual-loop structure (Fig.~\ref{fig_preliminary}(b)). This dataflow is well-suited for hardware acceleration and is often mapped onto efficient pipeline architectures (Fig.~\ref{fig_preliminary}(c)) composed of parallel MAC units~\cite{DaduRBD,liuDRACOCodesignDSPEfficient2025a}.

\subsection{Optimization Objective} \label{section_2_2}

Hardware accelerator design inherently involves trade-offs between performance, power, and area. For practical robotic systems, performance (i.e., throughput and latency) is not an optimization variable but a hard constraint dictated by the control frequency required for system stability~\cite{DaduRBD}. To meet this performance target, the accelerator must employ a specific degree of parallelism, which translates to a fixed number of MAC units and thus a solidified high-level architecture~\cite{liuDRACOCodesignDSPEfficient2025a}. As these computations are increasingly deployed on resource-constrained edge devices~\cite{Corki}, minimizing power and area becomes the paramount optimization objective.

Consequently, the design challenge becomes a constrained optimization problem: minimizing the hardware cost (area and power) of MAC units by optimizing their data bit-widths (i.e., quantizing the variables in the RBD algorithm) while satisfying the user-defined motion accuracy and hardware performance requirements. In this work, we focus on the fixed-point data format as it allows for efficient hardware implementation using integer arithmetic units, making it the standard choice in related works~\cite{Robomorphic,DaduRBD,liuDRACOCodesignDSPEfficient2025a}.

\subsection{Motivation} \label{section_2_3}

Prior work~\cite{liuDRACOCodesignDSPEfficient2025a} employs a uniform-precision quantization framework (Fig.~\ref{fig_preliminary}(d)), which applies a single fixed-point data format to all variables, meaning they share the same integer (determining range) and fractional (determining precision) bit-widths. This simplicity permits an exhaustive search, using time-consuming end-to-end simulations to evaluate the robot's trajectory error. However, this one-size-fits-all approach is inherently suboptimal. The core reason is that internal variables within RBD computations exhibit widely varying sensitivities to quantization; reducing their precision impacts the final motion accuracy to different extents.

This observation motivates a mixed-precision approach, where a tailored bit-width is assigned to each variable. Such a fine-grained strategy promises significant gains in area and power efficiency by allocating precision only where it is needed. However, breaking away from uniform precision transforms the design problem into a vast search space. For a space of this magnitude, the brute-force, simulation-based methods used in uniform-precision quantization become computationally intractable, necessitating the intelligent and efficient exploration methodology proposed in this work.

\section{Mixed-Precision Quantization Framework} \label{section_3}

\subsection{Application Workload Characterization} \label{section_3_1} 

To achieve optimal hardware efficiency, our framework generates a customized quantization scheme tailored to the specific robot and its target application. Consequently, this design-space exploration serves as an offline, one-time process for each new deployment scenario. The procedure initializes by constructing an application-centric profiling dataset. This dataset comprises representative motion workloads, such as trajectory-tracking paths for a manipulator~\cite{adabagMPCGPURealTimeNonlinear2024}, alongside stress tests like high-velocity maneuvers. This tailored benchmark directly drives all downstream stages, including sensitivity analysis and surrogate model training, to guarantee that the resulting hardware configuration is strictly optimized for its intended real-world operating conditions.

\subsection{Variable Grouping and Sensitivity Analysis} \label{section_3_2}

The vast number of variables within RBD computations creates a combinatorially explosive search space for quantization. To make this tractable, \textit{Stage 1} of our framework is dedicated to search space pruning (Fig.~\ref{fig_quantization_framwork}). We achieve this by introducing a methodology to group variables and analyze their quantization sensitivity. This approach reduces the search complexity by focusing on groups rather than individual parameters, while preserving precision for critical computations. Additionally, this stage utilizes module-level simulations to determine the dynamic range, thereby defining the required integer bit-width for each group to prevent overflow.

\subsubsection{Initial Grouping} 
\ 
\newline
The core RBD algorithms involve numerous variables (intermediate variables) that are candidates for quantization. To make the search space manageable, we first apply a heuristic grouping based on their physical and algorithmic roles. This step is performed offline at design-time as the underlying physical and algorithmic principles are fundamental to RBD and thus universally applicable~\cite{RBDAlgorithms}. This initial grouping reduces the number of entities to be quantized, shrinking the search space from over $12^{50}$ to a more tractable $12^{27}$.

\begin{figure}[t]
    \centering
    \includegraphics[width=70mm]{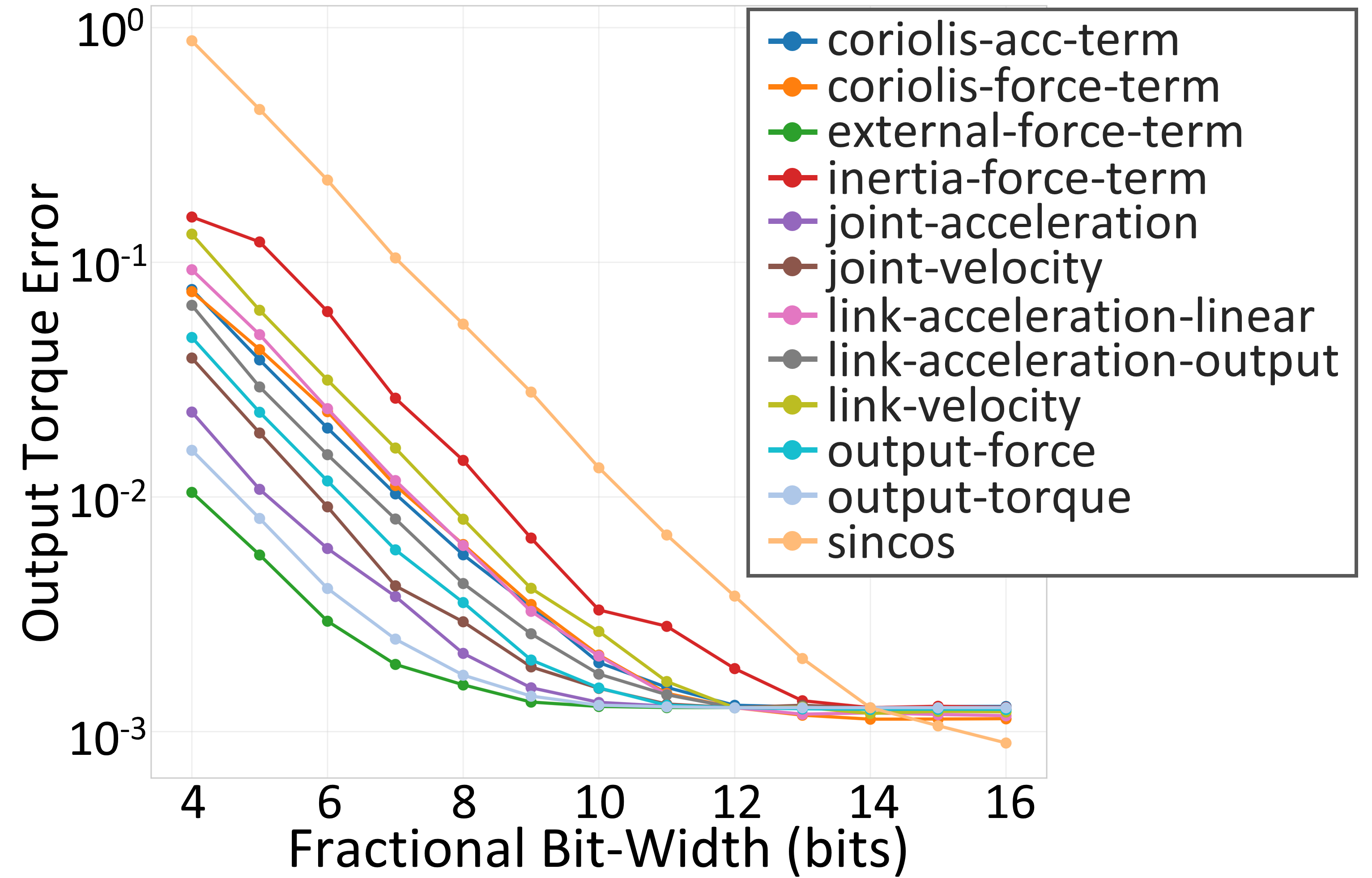}
    \vspace{-10pt}
    \caption{Sensitivity curves of all variable groups in RNEA.}
    \label{fig_sensitivity_curve}
\end{figure}

\subsubsection{Module-level Sensitivity Analysis} \label{section_3_2_2}
\ 
\newline
The next step is to quantify each group's quantization sensitivity, a foundational analysis for subsequent optimization. In this context, sensitivity measures how variations in a group's bit-width impact the robot's final motion trajectory. Accurately evaluating this impact requires computationally prohibitive end-to-end closed-loop simulations~\cite{liuDRACOCodesignDSPEfficient2025a}. To bypass this bottleneck, we perform a rapid, module-level analysis, utilizing a module's output error (e.g., torque error from RNEA) as a reliable proxy for the final trajectory error. 

A critical challenge in this analysis is error coupling: a group's sensitivity is not static, but rather fluctuates depending on the quantization states of all other groups. To account for this interplay, we avoid evaluating any group in a vacuum. Instead, we sweep a target group's fractional bit-width across its valid range while locking other groups into a fixed "background" configuration. To ensure robustness, we repeat this targeted sweep across multiple distinct background configurations systematically generated via Latin Hypercube Sampling~\cite{LHS}. This approach guarantees broad coverage of the high-dimensional parameter space, ultimately yielding a comprehensive family of error-versus-bit-width sensitivity curves for each group. Fig.~\ref{fig_sensitivity_curve} illustrates one such set of curves generated under a single background configuration.

To achieve scenario-adaptivity, we perform this sensitivity analysis on the specific workloads of the target application. As illustrated in Fig.~\ref{fig_quantization_framwork}, this stage yields three key outputs. First, we extract each group's sensitivity curves, which guide subsequent group merging, as well as local refinement heuristics (Sec.~\ref{section_3_4_3}, blue dotted arrow in Fig.~\ref{fig_quantization_framwork}). Second, the process generates module-level error datasets that serve as essential prior features for surrogate model training (Sec.~\ref{section_3_3}, orange dotted arrow in Fig.~\ref{fig_quantization_framwork}). Third, we determine each group's required integer bit-width to prevent numerical overflow. Because the integer portion dictates the representable range in fixed-point arithmetic, we derive this width by capturing the maximum dynamic range during module-level simulations and expanding it with a predefined safety margin to accommodate unseen runtime dynamics.

\subsubsection{Group Merging and Validation} 
\ 
\newline
To further simplify the problem, we merge variable groups that exhibit similar sensitivity profiles, quantified using metrics such as Mean Squared Error on their sensitivity curves. By constraining all variables within a newly merged group to share a uniform bit-width, this aggregation inevitably introduces slight precision degradation. Consequently, each potential merge must be rigorously validated; it is only accepted if the resulting module-level error remains below an empirically determined threshold~\cite{liuDRACOCodesignDSPEfficient2025a}, ensuring the merge does not create a new error bottleneck. This process also defines a viable fractional bit-width search range for each group by pruning any lower-precision configurations that exceed predefined module-level error thresholds. After group merging, the search space is further compressed to approximately $7^{13}$, with minor variations depending on the specific workload.

In summary, this stage transforms the problem by grouping and merging variables to reduce search dimensionality. It establishes the required integer bit-width for each group and prunes their fractional bit-width ranges. This creates a compact search problem and generates essential data-driven priors for the subsequent stages. 

\subsection{Surrogate Model for Trajectory Error Prediction} \label{section_3_3}

As analyzed in Sec.~\ref{section_1}, the prohibitive computational cost of end-to-end closed-loop simulations constitutes the primary bottleneck for extensive design space exploration. To bypass this limitation, we introduce an efficient surrogate model designed to map the quantized bit-width configurations across all groups directly to the final trajectory error. By replacing costly physical simulations with rapid algorithmic predictions, this surrogate forms the cornerstone of our hardware-aware search strategy.

\subsubsection{Model Selection} 
\ 
\newline
A universal surrogate model is impractical, as the characteristics of trajectory error are highly workload-dependent. Specifically, the mechanisms of error propagation and dynamic responses vary drastically across different robotic platforms (e.g., varying degrees of freedom, joint types, and link morphologies) and diverse application scenarios (e.g., high-speed locomotion versus high-precision manipulation)~\cite{Robomorphic,machines13090771}. 
Consequently, training a "one-size-fits-all" model would require a large dataset encompassing the combinatorial explosion of all possible robots and tasks. However, acquiring such comprehensive, high-fidelity robotic data is notoriously difficult and resource-intensive in the robotics community~\cite{brohan2023open,dulac2019challenges}. Furthermore, even if such an exhaustive dataset could be curated, the continuous emergence of novel robotic designs and operational environments would quickly render a static generalized model obsolete. Therefore, in practice, the surrogate model must be workload-specific---trained uniquely at runtime for each new scenario.

Establishing a scenario-specific surrogate model requires ground-truth training data, consisting of mixed-precision configurations and their corresponding trajectory errors. However, acquiring each sample necessitates full end-to-end closed-loop simulations, a process that is highly computationally expensive. While this exploration represents a one-time cost for a finalized application, the robotic hardware-software co-design process~\cite{sacksRoboXEndtoEndSolution2018,Corki} is inherently iterative. System specifications, such as workload definitions, control parameters, and robot kinematics, frequently evolve during development. Consequently, allocating thousands of simulation hours to collect extensive training datasets would bottleneck this rapid prototyping cycle. Constrained by a practical time budget, the framework can only afford to evaluate a minimal number of configurations, inevitably resulting in severe data scarcity for surrogate model training.

Such scarcity renders data-hungry models like DNNs impractical, establishing our first requirement for extreme data efficiency. Furthermore, because the optimization engine must query the surrogate model thousands of times to evaluate candidate configurations during the exploration phase, the model must provide low inference latency to maintain overall search efficiency. These combined constraints define our core requirement: an accurate model that is both \textbf{data-efficient} and \textbf{computationally lightweight}. Our goal, therefore, is not to achieve global predictive perfection across all domains, but to deploy a model that serves as a data-efficient, locally-accurate guide for the optimizer within the target workload.

For this specific priority on local guidance under data scarcity, Random Forest (RF)~\cite{grinsztajn2022why} emerges as an ideal choice. While any model trained on a small dataset risks overfitting, RF's ensemble nature provides inherent robustness against this issue~\cite{breiman2001random}. To further mitigate overfitting on our deliberately small training set, we strictly regularize the RF hyperparameters, limiting the maximum tree depth to 8$\sim$10 and enforcing a minimum of 3$\sim$5 samples per leaf node. Furthermore, its ensemble of decision trees is empirically proven to be highly effective on limited tabular data, generally exhibiting superior performance compared to deep learning models on small datasets~\cite{grinsztajn2022why}. This structure is particularly adept at capturing the complex non-linearities and high-order interaction effects that link bit-width assignments to final trajectory error.

\subsubsection{Hierarchical Feature Engineering}
\ 
\newline
\textbf{Hierarchical Framework.} A direct mapping from bit-width assignments to final trajectory error presents a significant learning challenge. This relationship is obscured by complex, non-linear error propagation through the robotic control stack~\cite{liuDRACOCodesignDSPEfficient2025a}. To alleviate this, rather than treating the entire control loop as a complex, monolithic end-to-end prediction task, our approach decomposes the learning task into two modular sub-stages. In the RBD computation flow, computations are executed sequentially through distinct core modules~\cite{DaduRBD}. For instance, the RNEA module calculates the required joint torques~\cite{AnalyticalDerivativesRSS}, while the Minv module computes the inverse of the joint-space mass matrix~\cite{carpentierAnalyticalInverseJoint}. Therefore, our first sub-stage constructs an array of independent predictors. Each predictor explicitly models the output error of a specific core RBD module (e.g., the torque error generated by the RNEA) based exclusively on its local bit-width assignments. These intermediate, physically meaningful error predictions then serve as the principal features for the second sub-stage, which evaluates their combined impact on the final trajectory error. This module-guided decomposition allows the framework to learn from crucial intermediate states rather than attempting to infer them from raw bit-widths alone.

\textit{Sub-stage 1: Module-Level Error Characterization.}
We train a set of lightweight RF models dedicated to individual core RBD modules, utilizing the error datasets generated during the sensitivity analysis of \textit{Stage 1} (Fig.~\ref{fig_quantization_framwork}). Instead of targeting the final trajectory, these independent predictors exclusively map bit-width assignments to localized output errors (e.g., $e_{RNEA}$, $e_{Minv}$).

\textit{Sub-stage 2: System-Level Error Prediction.} 
The second sub-stage predicts the final trajectory error. Moving beyond raw bit-widths, we construct a comprehensive, physically-grounded feature space for this final predictor by integrating the outputs from Sub-stage 1 with domain-specific heuristics:

\ding{202} Module-Level Error Priors: The localized output errors predicted in Sub-stage 1 are directly incorporated as foundational prior features to explicitly represent the quantization impact within each individual module.

\ding{203} Derived Error Metrics: We engineer features from the Sub-stage 1 priors to capture the controller's response to relative inaccuracies between dynamic components. For example, we utilize the module error ratio ($e_{RNEA} / e_{Minv}$) and the normalized error imbalance ($|e_{RNEA} - e_{Minv}| / (e_{RNEA} + e_{Minv})$) to quantify the distribution and severity of these intermediate inaccuracies.

\ding{204} Domain-Specific Dependencies: We incorporate structural knowledge from robotics and control theory. These features abstract key concepts, such as computational bottlenecks and precision imbalances, providing insights difficult to learn from raw bit-widths alone. For instance, in the RNEA forward pass, where joint $i$'s velocity depends on joint $i-1$'s~\cite{liuDRACOCodesignDSPEfficient2025a}, we create a feature representing the minimum bit-width within this recursive chain. This reflects the control principle that a cascaded system's accuracy is limited by its weakest link~\cite{RBDAlgorithms}. 

\ding{205} Base and Interaction Terms: Finally, the bit-width of each group and their pairwise products are included to capture any remaining linear and second-order quantization effects.

\textbf{Feature Selection.} To mitigate overfitting within the resulting high-dimensional feature space, we embed feature selection directly into the model training phase. Rather than employing isolated preprocessing filters~\cite{guyon2003introduction}, we leverage the intrinsic feature importance mechanism of RF~\cite{louppe2013understanding}. This tree-based approach evaluates the quantitative importance of each candidate based on its overall contribution to error reduction, which is particularly well-suited for our methodology: it naturally captures the complex, non-linear interactions among variables that simple linear methods (e.g., Pearson correlation) fail to detect~\cite{hastie2009elements}. Furthermore, its scale-invariance ensures an unbiased evaluation across features exhibiting vastly different dynamic ranges.

We rank all feature candidates by importance scores, selecting the top 20 most salient features to train the final surrogate model.

\begin{algorithm}[t]
\begin{spacing}{0.9}
\caption{Cost-Driven Greedy Refinement}
\label{alg:cost_greedy_revised}
\begin{algorithmic}[1]
\State \textbf{Input:} Candidate Pool $P$, Hardware Impact Map $H$
\State \textbf{Output:} Candidate Pool $P$
\State $G_{sorted} \gets$ Sort groups $g$ based on descending $H[g]$
\For{each configuration $B \in P$}
    \For{each group $g_i \in G_{sorted}$}
        \State $B' \gets \text{ReduceBitwidth}(B, g_i)$
        \If{$Error_{pred}(B') > 1.1 \cdot \epsilon_{\text{max}}$}
            \State $B' \gets \text{Compensate}(B', G_{sorted}[i+1:])$
        \EndIf
        \State $\text{UpdatePoolIfBetter}(P, B')$
    \EndFor
\EndFor
\end{algorithmic}
\end{spacing}
\end{algorithm}

\subsection{Hardware-Aware Searching} \label{section_3_4}

With the surrogate model for trajectory error prediction, our objective is to find the optimal fractional bit-width assignments that minimize hardware cost while satisfying a user-defined error tolerance. To solve this constrained optimization problem, we employ a hybrid search strategy that synergizes global exploration with local refinement, guided by a fast analytical cost estimator.

\subsubsection{Analytical Hardware Cost Estimator} 
\ 
\newline
To guide the search, we develop a fast, cell-library-based analytical model that maps a quantization configuration to a hardware cost (area and power) estimate, grounded in a state-of-the-art (SOTA) RBD accelerator architecture~\cite{liuDRACOCodesignDSPEfficient2025a}. The model's logic is straightforward: for a given bit-width assignment $B$, it first counts the required instances ($N_u$) of each unique arithmetic operator $u$ from the set of all operator instances $U$, where uniqueness is defined by the operator's function and operand bit-widths (e.g., a 12$\times$8-bit multiplier). The cost for each operator type, $C(u)$, is pre-characterized via a one-time synthesis of multipliers and adders with various bit-widths, and stored in a look-up table (LUT). The total hardware cost is then computed by summing the costs of all operator instances:
\begin{equation}
\text{Cost}(B) = \sum_{u \in U} N_u \cdot C(u)~.
\end{equation}
This approach focuses on arithmetic logic. While quantization also affects FIFOs, control logic, and routing congestion, our post-layout analysis confirms these secondary effects are negligible compared to the cost variation of arithmetic units.

\subsubsection{Bayesian Optimization} \label{section_3_4_2}  
\ 
\newline
For global exploration, we employ Bayesian Optimization (BO), a sample-efficient method ideal for finding the global optimum of a black-box function under complex constraints~\cite{tu2025rsizing,ma2019cad}. Guided by the analytical cost estimator and the surrogate trajectory error predictor (Sec.~\ref{section_3_3}), BO is tasked with finding the bit-width configuration $B$ that solves the following optimization problem:
$$
\begin{aligned}
& \underset{B}{\text{minimize}}
& & \text{Cost}(B) = \text{Area}(B) + \lambda \cdot \text{Power}(B) \\
& \text{subject to}
& & \text{Error}_{\text{predict}}(B) \le 1.1 \cdot \epsilon_{\text{max}}
\end{aligned}
$$
The objective function, $\text{Cost}(B)$, is the weighted hardware cost from our analytical estimator. The factor $\lambda$ is used to balance the trade-off between area and power. The surrogate's prediction must satisfy the user-defined tolerance ($\epsilon_{\text{max}}$). We relax this boundary by 10\% to offset potential surrogate underestimation, empirically balancing solution feasibility and search efficiency. The search concludes by identifying the top 30 configurations that satisfy the constraint, creating a candidate pool for final validation.

\begin{algorithm}[t]
\begin{spacing}{0.9}
\caption{Sensitivity-Guided Tuning}
\label{alg:sensitivity_tuning_revised}
\begin{algorithmic}[1]
\State \textbf{Input:} Candidate Pool $P$, Sensitivity Map $S$
\State \textbf{Output:} Candidate Pool $P$
\State $G_{high}, G_{low}, G_{other} \gets$ Partition groups based on $S[g]$
\For{each configuration $B \in P$}
    \State $B' \gets \text{ReduceBitwidth}(B, G_{low})$
    \State $B' \gets \text{IncreaseBitwidth}(B', G_{high})$
    \If{$Error_{pred}(B') \le 1.1 \cdot \epsilon_{\text{max}}$}
        \State $B'' \gets \text{ReduceBitwidth}(B', G_{other})$
        \State $\text{UpdatePoolIfBetter}(P, B'')$
    \EndIf
\EndFor
\end{algorithmic}
\end{spacing}
\end{algorithm}

\subsubsection{Local Refinements} \label{section_3_4_3}
\ 
\newline
While BO effectively identifies promising regions, it is not designed for finding precise local minima~\cite{eriksson2019scalable}. We therefore sequentially apply two complementary local refinement heuristics. These greedy-based algorithms are computationally efficient and leverage domain-specific knowledge for targeted fine-tuning where BO cannot.

The first, Cost-Driven Greedy Refinement (Alg.~\ref{alg:cost_greedy_revised}), takes a cost-first approach. It ranks variable groups in descending order of their hardware impact (i.e., the aggregate hardware cost of the arithmetic units they affect). Traversing this sorted list, the algorithm greedily reduces the bit-width of high-impact groups to maximize hardware savings. If reducing the bit-width of the current group causes the trajectory error to exceed the predefined threshold, the framework initiates a compensation mechanism. To recover system accuracy, this step selectively allocates higher bit-widths to the remaining, less sensitive groups within the processing queue.

The second, Sensitivity-Guided Tuning, shown as Alg.~\ref{alg:sensitivity_tuning_revised}, adopts an accuracy-first strategy by leveraging the sensitivity analysis from Sec.~\ref{section_3_2_2}. To systematically adjust the configurations, the algorithm first partitions all variable groups into three classes: high-sensitivity, low-sensitivity, and others. It then initiates a precision trade-off by reducing the bit-widths of low-sensitivity groups, then boosting the high-sensitivity ones. Building on this trade-off, if the configuration still meets the required error bound, the algorithm then proceeds to cautiously reduce the bit-widths of the remaining "other" groups to further maximize hardware efficiency.

In both heuristics, any newly discovered configuration with a lower hardware cost updates the candidate pool, effectively exploring the local neighborhood for a better solution.

\subsubsection{End-to-End Validation} 
\ 
\newline
The final step is to validate the pool of 30 candidate configurations produced by our search framework. Each candidate undergoes a rigorous end-to-end simulation to determine its true trajectory error. From the candidates in this pool that verifiably satisfy the user-defined tolerance, we select the one with the minimum hardware cost as the final solution.

\begin{figure}[t]
    \centering
    \includegraphics[width=62mm]{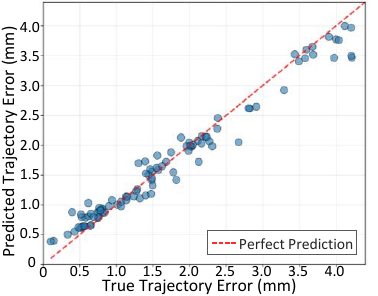}
    \vspace{-12pt}
    \caption{Predicted versus true trajectory error.}
    \label{fig_prediction_vs_true}
    \vspace{-8pt}
\end{figure}

\section{Evaluation} \label{section_4}

\subsection{Experimental Setup} \label{section_4_1} 

To demonstrate the generalizability and scalability of our framework, we conduct comprehensive evaluations across three diverse robotic platforms: the 7-Degree-of-Freedom (DoF) iiwa manipulator~\cite{iiwa}, the 12-DoF HyQ quadruped~\cite{HyQDynamicLegged}, and the 29-DoF Atlas humanoid~\cite{Atlas}. For each robot, we curate a domain-specific application workload comprising 40 representative trajectories. Specifically, the iiwa workload covers trajectory tracking and common manipulation tasks~\cite{Corki,adabagMPCGPURealTimeNonlinear2024}; the HyQ workload features locomotion profiles~\cite{dicarloDynamicLocomotionMIT2018,MITCheetah} such as walking, trotting, and climbing; and the Atlas workload encompasses complex whole-body motions~\cite{Optimizationbasedlocomotion,dantecWholeBodyModelPredictive2022}, including walking, squatting, and dynamic balancing. These trajectory sets serve as our primary test cases; notably, the framework's search time scales proportionally with the workload size, ensuring tractability for even more extensive applications. The control and motion simulation environment utilizes the Pinocchio library~\cite{PinocchioLibrary} and a model predictive controller~\cite{adabagMPCGPURealTimeNonlinear2024}, mirroring the setup in DRACO~\cite{liuDRACOCodesignDSPEfficient2025a} to ensure a fair comparison. Finally, all hardware metrics are derived from post-layout results using the Synopsys IC Compiler under the TSMC 28-nm CMOS technology library.

\vspace{-2pt}
\subsection{Surrogate Model Effectiveness} \label{section_4_2} 

We evaluate the effectiveness of our trajectory error predictor and quantify the benefit of feature engineering. Given that data collection via end-to-end simulation is the dominant cost in the exploration process, the predictor (surrogate model) is trained on a deliberately small training dataset of 100 samples. Specifically, each sample represents a distinct mixed-precision bit-width configuration, and its corresponding label is the real trajectory error evaluated through end-to-end closed-loop simulation. As plotted in Fig.~\ref{fig_prediction_vs_true}, the predicted versus true trajectory errors on a separate 80-sample test set exhibit a strong linear correlation. Quantitatively, the model achieves excellent predictive accuracy, evidenced by a high coefficient of determination ($R^2$) of 0.937 and a low Root Mean Square Error (RMSE) of 0.307 (Table~\ref{tab:ablation_study}). More critically for the optimization objective, where correct ranking is paramount, the model exhibits strong rank fidelity, achieving a Spearman's correlation of 0.95 and a Kendall's Tau of 0.81. This confirms its capability to guide the optimization process by correctly ordering design candidates. Conversely, models relying exclusively on raw bit-width features converge to suboptimal design points, incurring an average area overhead of 12\% and a power penalty of 10\%.

\begin{figure}[t]
    \centering
    \includegraphics[width=66mm]{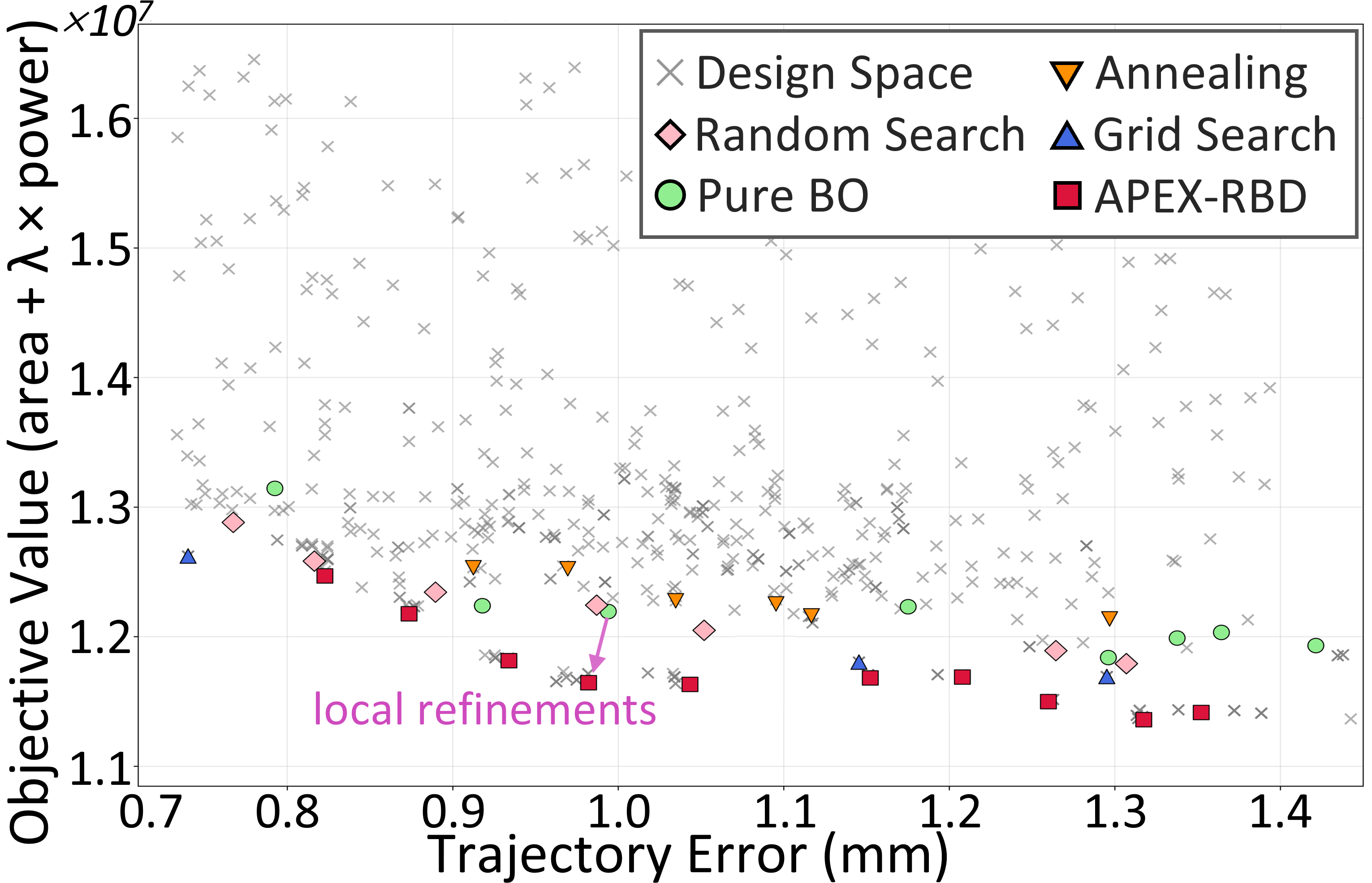}
    \vspace{-10pt}
    \caption{Comparison of search results under different trajectory error constraints with proposed APEX-RBD and various search algorithms (random search, pure Bayesian Optimization, annealing algorithm, and grid search).}
    \label{fig_search_results}
\end{figure}

\begin{table}[t]\small 
    \caption{Ablation Study on Feature Engineering.}
    \centering
    \vspace{-7pt}
    \label{tab:ablation_study}
    \renewcommand\arraystretch{1.06} 
    \begin{tabular}{m{2.7cm}<{\centering} m{0.8cm}<{\centering} m{1.1cm}<{\centering} m{1cm}<{\centering} m{1cm}<{\centering}}
    \hline
    Feature Set & \textbf{$R^2$} & RMSE & Spearman & Kendall \\ \hline
    Bit-widths only & 0.879 & 0.398 & 0.89 & 0.71 \\ 
    + Module-Level Priors     & 0.922 & 0.332 & 0.93 & 0.79 \\ 
    All Features   & \textbf{0.937} & \textbf{0.307} & \textbf{0.95} & \textbf{0.81} \\ \hline
    \end{tabular}
\end{table}

\vspace{-2pt}
\subsection{Search Strategy Performance} \label{section_4_3} 

We evaluate the effectiveness of our search strategy by comparing APEX-RBD against other baseline algorithms~\cite{kwonRLPTQRLbasedMixed2024}. The optimization objective is to minimize the hardware cost (Sec.~\ref{section_3_4_2}), where the weighting factor $\lambda$ is set to $10^4$ to normalize the different orders of magnitude between area ($\mu\text{m}^2$) and power (mW). To ensure a fair assessment, all compared algorithms utilize the same surrogate model and hardware cost estimator in our framework.

As illustrated in Fig.~\ref{fig_search_results}, APEX-RBD discovers superior configurations that define a performance envelope of lower hardware cost for any given error threshold. While pure BO identifies promising regions, our proposed local refinements (Sec.~\ref{section_3_4_3}) provide a significant advantage. This is visually confirmed near the 1.0 mm error threshold, where APEX-RBD finds a candidate with lower cost than pure BO (pink arrow). This demonstrates that our heuristics effectively conduct fine-grained exploration around the configurations identified by BO to uncover superior designs, creating a powerful synergy between global exploration and local search.

\begin{figure*}[t]
    \centering
    \includegraphics[width=\linewidth]{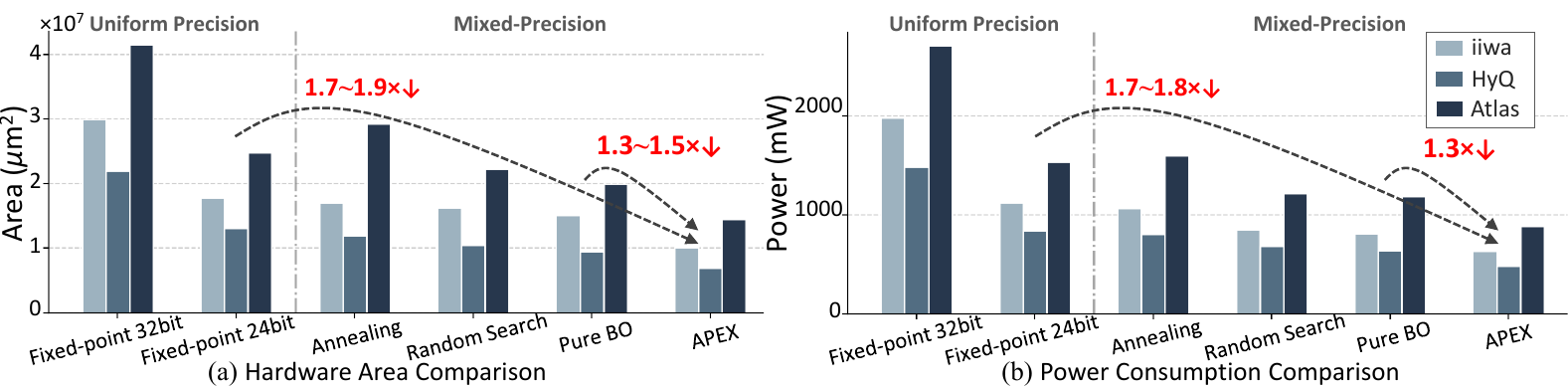}
    \vspace{-22pt}
    \caption{Hardware area and power cost comparison through different quantization methods, using Synopsys IC Compiler based on TSMC 28-nm CMOS technology.}
    \label{fig_hardware_comparison}
    \vspace{-9pt}
\end{figure*}

\vspace{-2pt}
\subsection{End-to-End Hardware Gains} \label{section_4_4} 

We evaluate the end-to-end hardware gains of our framework on iiwa, HyQ, and Atlas robots, to align with the evaluation setups of existing robotic works~\cite{Roboshape,DaduRBD,liuDRACOCodesignDSPEfficient2025a}. To ensure a fair and comprehensive comparison, we benchmark APEX-RBD against two categories of baselines. The first category comprises uniform-precision baselines, specifically 32-bit~\cite{DaduRBD} and 24-bit~\cite{liuDRACOCodesignDSPEfficient2025a}, which represent the configurations of SOTA RBD accelerators~\cite{DaduRBD,liuDRACOCodesignDSPEfficient2025a}. The second category evaluates the search efficacy within the mixed-precision space. Since we propose the first mixed-precision quantization framework for this domain, we compare APEX-RBD against hardware designs derived from other common search algorithms, including Simulated Annealing, Random Search, and Pure BO. Notably, Pure BO represents the search outcome without the two local refinement algorithms proposed in Sec.~\ref{section_3_4_3}. All comparisons are conducted under an identical throughput constraint, which is enforced by matching the number of parallel MAC units. 

To facilitate this hardware evaluation, we develop a compiler to generate RTL from any given bit-width configuration, based on the architecture of~\cite{liuDRACOCodesignDSPEfficient2025a}. The resulting hardware costs are then derived from post-layout results. Furthermore, the motion accuracy of all mixed-precision configurations evaluated is constrained to be on par with the uniform 24-bit baseline. Prior works~\cite{Corki,liuDRACOCodesignDSPEfficient2025a} have confirmed that this precision level has a negligible impact on task success rates, thereby preserving overall system stability.


\begin{figure}[t]
    \centering
    \includegraphics[width=70mm]{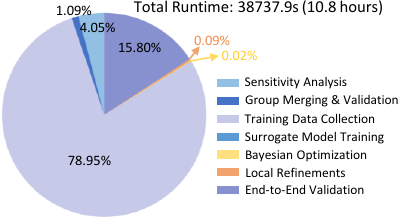}
    \vspace{-10pt}
    \caption{Runtime breakdown of APEX-RBD for iiwa.}
    \label{fig_time_breakdown}
    \vspace{-12pt}
\end{figure}

As illustrated in Fig.~\ref{fig_hardware_comparison}, APEX-RBD achieves up to 1.9$\times$ area reduction and 1.8$\times$ power savings compared to the 24-bit uniform baseline, while delivering the same motion accuracy. Moreover, within the mixed-precision category, APEX-RBD consistently discovers more area- and power-efficient designs than other search algorithms. Specifically, compared to the configuration found by Pure BO, APEX-RBD achieves up to 1.5$\times$ area and 1.3$\times$ power savings. This performance gap highlights the effectiveness of the local refinement heuristics for escaping local optima. Finally, the hardware benefits are more pronounced when compared against the conservative 32-bit baseline, where APEX-RBD realizes up to 3.1$\times$ savings in both area and power.

\vspace{-2pt}
\subsection{Framework Runtime Analysis} \label{section_4_5} 

We evaluate APEX-RBD's runtime by performing mixed-precision exploration for an iiwa robot. For a given application, this exploration is a practical one-time cost. In our evaluation, the entire process completed in 10.8 hours on a single commodity workstation (Intel i9 CPU, 32GB RAM). This result is highly competitive compared to DRACO~\cite{liuDRACOCodesignDSPEfficient2025a}, which requires 8 hours to search for a much simpler uniform-precision configuration. As detailed in Fig.~\ref{fig_time_breakdown}, the process is dominated by simulation-based data collection (78.95\%), making data acquisition the primary bottleneck. The overall efficiency of APEX-RBD stems from two key strategies.

First, the group merging stage drastically reduces the problem's complexity by shrinking the search space. This step has a negligible runtime (1.09\%) but is essential as it directly reduces the required training samples for surrogate model. Second, our use of a data-efficient RF surrogate replaces costly simulations and requires only 100 samples. This choice stands in stark contrast to a DNN-based alternative, which our experiments show would need over 300 samples to achieve comparable accuracy, extending the search by 18 hours. Together, these strategies make the dominant data collection phase tractable and enable the framework's overall efficiency.

\section{Conclusion}

We propose APEX-RBD, the first framework to automate mixed-precision quantization for RBD computations. It makes the search space tractable by integrating physics-aware space pruning with a surrogate-guided hybrid optimizer. APEX-RBD discovers designs achieving up to 1.9$\times$ area and 1.8$\times$ power reductions over uniform-precision baselines while satisfying accuracy constraints.

\vspace{-3pt}
\begin{acks}
This work was partially supported by the CRF C5032-23G and ACCESS — AI Chip Center for Emerging Smart Systems, the InnoHK initiative of ITC.
\end{acks}

\bibliographystyle{ACM-Reference-Format}
\bibliography{refs}

@article{carpentierAnalyticalInverseJoint,
  title = {Analytical {{Inverse}} of the {{Joint Space Inertia Matrix}}},
  author = {Carpentier, Justin},
  year = {2020}
}

@inproceedings{PinocchioLibrary,
  title = {The {{Pinocchio C}}++ Library : {{A}} Fast and Flexible Implementation of Rigid Body Dynamics Algorithms and Their Analytical Derivatives},
  booktitle = {2019 {{IEEE}}/{{SICE International Symposium}} on {{System Integration}} ({{SII}})},
  author = {Carpentier, Justin and Saurel, Guilhem and Buondonno, Gabriele and Mirabel, Joseph and Lamiraux, Florent and Stasse, Olivier and Mansard, Nicolas},
  year = {2019}
}

@inproceedings{AnalyticalDerivativesRSS,
  title = {Analytical Derivatives of Rigid Body Dynamics Algorithms},
  booktitle = {Robotics: Science and Systems},
  author = {Carpentier, Justin and Mansard, Nicolas},
  year = {2018}
}

@inproceedings{Corki,
  title = {Dadu-{{Corki}}: {{Algorithm-Architecture Co-Design}} for {{Embodied AI-powered Robotic Manipulation}}},
  booktitle = {Proceedings of the 52nd {{Annual International Symposium}} on {{Computer Architecture}}},
  author = {Huang, Yiyang and Hao, Yuhui and Yu, Bo and Yan, Feng and Yang, Yuxin and Min, Feng and Han, Yinhe and Ma, Lin and Liu, Shaoshan and Liu, Qiang and Gan, Yiming},
  year = {2025}
}

@inproceedings{Robomorphic,
  title = {Robomorphic Computing: A Design Methodology for Domain-Specific Accelerators Parameterized by Robot Morphology},
  booktitle = {Proceedings of the 26th {{ACM International Conference}} on {{Architectural Support}} for {{Programming Languages}} and {{Operating Systems}}},
  author = {Neuman, Sabrina M. and Plancher, Brian and Bourgeat, Thomas and Tambe, Thierry and Devadas, Srinivas and Reddi, Vijay Janapa},
  year = {2021}
}

@inproceedings{Roboshape,
  title = {{{RoboShape}}: {{Using Topology Patterns}} to {{Scalably}} and {{Flexibly Deploy Accelerators Across Robots}}},
  booktitle = {Proceedings of the 50th {{Annual International Symposium}} on {{Computer Architecture}}},
  author = {Neuman, Sabrina M. and Ghosal, Radhika and Bourgeat, Thomas and Plancher, Brian and Reddi, Vijay Janapa},
  year = {2023}
}

@inproceedings{plancherGRiDGPUAcceleratedRigid2022,
  title = {{{GRiD}}: {{GPU-Accelerated Rigid Body Dynamics}} with {{Analytical Gradients}}},
  booktitle = {2022 {{International Conference}} on {{Robotics}} and {{Automation}} ({{ICRA}})},
  author = {Plancher, Brian and Neuman, Sabrina M. and Ghosal, Radhika and Kuindersma, Scott and Reddi, Vijay Janapa},
  year = {2022}
}

@incollection{plancherPerformanceAnalysisParallel2020,
  title = {A {{Performance Analysis}} of {{Parallel Differential Dynamic Programming}} on a {{GPU}}},
  booktitle = {Algorithmic {{Foundations}} of {{Robotics XIII}}},
  author = {Plancher, Brian and Kuindersma, Scott},
  editor = {Morales, Marco and Tapia, Lydia and {S{\'a}nchez-Ante}, Gildardo and Hutchinson, Seth},
  year = {2020},
  volume = {14}
}

@inproceedings{sacksRoboXEndtoEndSolution2018,
  title = {{{RoboX}}: {{An End-to-End Solution}} to {{Accelerate Autonomous Control}} in {{Robotics}}},
  booktitle = {2018 {{ACM}}/{{IEEE}} 45th {{Annual International Symposium}} on {{Computer Architecture}} ({{ISCA}})},
  author = {Sacks, Jacob and Mahajan, Divya and Lawson, Richard C. and Khaleghi, Behnam and Esmaeilzadeh, Hadi},
  year = {2018}
}

@inproceedings{wanRoboticComputingFPGAs2022,
  title = {Robotic {{Computing}} on {{FPGAs}}: {{Current Progress}}, {{Research Challenges}}, and {{Opportunities}}},
  booktitle = {2022 {{IEEE}} 4th {{International Conference}} on {{Artificial Intelligence Circuits}} and {{Systems}} ({{AICAS}})},
  author = {Wan, Zishen and Lele, Ashwin and Yu, Bo and Liu, Shaoshan and Wang, Yu and Reddi, Vijay Janapa and Hao, Cong and Raychowdhury, Arijit},
  year = {2022}
}

@inproceedings{DaduRBD,
  title = {Dadu-{{RBD}}: {{Robot Rigid Body Dynamics Accelerator}} with {{Multifunctional Pipelines}}},
  booktitle = {Proceedings of the 56th {{Annual IEEE}}/{{ACM International Symposium}} on {{Microarchitecture}} {(MICRO)} '23},
  author = {Yang, Yuxin and Chen, Xiaoming and Han, Yinhe},
  year = {2023},
}

@online{iiwa,
  title = {KUKA iiwa},
  url = {https://www.kuka.com/en-de/products/robot-systems/industrial-robots/lbr-iiwa},
  year = {2014}
}

@online{HyQDynamicLegged,
  title = {{{HyQ}} - {{Dynamic Legged Systems}} - {{IIT}}},
  url = {https://dls.iit.it/hyq},
  year = {2013}
}

@online{Atlas,
    url = {https://bostondynamics.com/atlas/},
    title = {Atlas},
    year = {2021}
}

@online{DSP48,
    url = {https://docs.amd.com/r/2021.2-English/ug1483-model-composer-sys-gen-user-guide/DSP48E},
    title = {Vitis Model Composer User Guide-DSP48E},
    year = {2021}
}

@book{RBDAlgorithms,
  title = {Rigid Body Dynamics Algorithms},
  author = {Roy, Featherstone},
  year = {2008}
}

@inproceedings{kwonRLPTQRLbasedMixed2024,
  title = {{{RL-PTQ}}: {{RL-based Mixed Precision Quantization}} for {{Hybrid Vision Transformers}}},
  booktitle = {Proceedings of the 61st {{ACM}}/{{IEEE Design Automation Conference}}},
  author = {Kwon, Eunji and Zhou, Minxuan and Xu, Weihong and Rosing, Tajana and Kang, Seokhyeong},
  year = {2024}
}

@article{paniegoModelOptimizationDeep2024,
  title = {Model {{Optimization}} in {{Deep Learning Based Robot Control}} for {{Autonomous Driving}}},
  author = {Paniego, Sergio and Paliwal, Nikhil and Ca{\~n}as, Jos{\'e}Mar{\'i}a},
  year = {2024},
  journal = {IEEE Robotics and Automation Letters},
  volume = {9},
  number = {1}
}

@inproceedings{parkSaliencyAwareQuantizedImitation2025,
  title = {Saliency-{{Aware Quantized Imitation Learning}} for {{Efficient Robotic Control}}},
  booktitle = {Proceedings of the {{IEEE}}/{{CVF International Conference}} on {{Computer Vision}}},
  author = {Park, Seongmin and Kim, Hyungmin and Kim, Sangwoo and Jeon, Wonseok and Yang, Juyoung and Jeon, Byeongwook and Oh, Yoonseon and Choi, Jungwook},
  year = {2025}
}

@inproceedings{ExplainableFuzzy,
author = {Fan, Hanwei and Wang, Ya and Li, Sicheng and Liang, Tingyuan and Zhang, Wei},
title = {Explainable Fuzzy Neural Network with Multi-Fidelity Reinforcement Learning for Micro-Architecture Design Space Exploration},
year = {2024},
booktitle = {Proceedings of the 61st ACM/IEEE Design Automation Conference},
series = {DAC '24}
}

@ARTICLE{LHS,
  author={M. D. McKay and R. J. Beckman and W. J. Conover},
  journal={Technometrics}, 
  title={A Comparison of Three Methods for Selecting Values of Input Variables in the Analysis of Output from a Computer Code}, 
  year={1979},
  volume={21},
  number={2}
}

@INPROCEEDINGS{liuDRACOCodesignDSPEfficient2025a,
  author={Liu, Xingyu and Liang, Jiawei and Zhang, Yipu and Du, Linfeng and Ma, Chaofang and Yu, Hui and Xu, Jiang and Zhang, Wei},
  booktitle={2026 IEEE International Symposium on High Performance Computer Architecture (HPCA)}, 
  title={DRACO: A Hardware-Efficient Robot Rigid Body Dynamics Accelerator with Precision-Aware Quantization Framework}, 
  year={2026}
}

@inproceedings{adabagMPCGPURealTimeNonlinear2024,
  title = {{{MPCGPU}}: {{Real-Time Nonlinear Model Predictive Control}} through {{Preconditioned Conjugate Gradient}} on the {{GPU}}},
  booktitle = {2024 {{IEEE International Conference}} on {{Robotics}} and {{Automation}} ({{ICRA}})},
  author = {Adabag, Emre and Atal, Miloni and Gerard, William and Plancher, Brian},
  year = 2024
}

@inproceedings{dicarloDynamicLocomotionMIT2018,
  title = {Dynamic {{Locomotion}} in the {{MIT Cheetah}} 3 {{Through Convex Model-Predictive Control}}},
  booktitle = {2018 {{IEEE}}/{{RSJ International Conference}} on {{Intelligent Robots}} and {{Systems}} ({{IROS}})},
  author = {Di Carlo, Jared and Wensing, Patrick M. and Katz, Benjamin and Bledt, Gerardo and Kim, Sangbae},
  year = 2018
}

@INPROCEEDINGS{MITCheetah,
  author={Bledt, Gerardo and Powell, Matthew J. and Katz, Benjamin and Di Carlo, Jared and Wensing, Patrick M. and Kim, Sangbae},
  booktitle={2018 IEEE/RSJ International Conference on Intelligent Robots and Systems (IROS)}, 
  title={MIT Cheetah 3: Design and Control of a Robust, Dynamic Quadruped Robot}, 
  year={2018}
  }

@article{Optimizationbasedlocomotion,
author = {Kuindersma, Scott and Deits, Robin and Fallon, Maurice and Valenzuela, Andr\'{e}s and Dai, Hongkai and Permenter, Frank and Koolen, Twan and Marion, Pat and Tedrake, Russ},
title = {Optimization-based locomotion planning, estimation, and control design for the atlas humanoid robot},
year = {2016},
journal = {Auton. Robots}
}

@inproceedings{dantecWholeBodyModelPredictive2022,
  title = {Whole-{{Body Model Predictive Control}} for {{Biped Locomotion}} on a {{Torque-Controlled Humanoid Robot}}},
  booktitle = {2022 {{IEEE-RAS}} 21st {{International Conference}} on {{Humanoid Robots}} ({{Humanoids}})},
  author = {Dantec, Ewen and Naveau, Maximilien and Fernbach, Pierre and Villa, Nahuel and Saurel, Guilhem and Stasse, Olivier and Taix, Michel and Mansard, Nicolas},
  year = 2022
}

@Article{machines13090771,
AUTHOR = {Yang, Shengnan and Jiang, Wenping and Long, Lin},
TITLE = {Data-Driven Method for Robotic Trajectory Error Prediction and Compensation Based on Digital Twin},
JOURNAL = {Machines},
YEAR = {2025}
}

@article{breiman2001random,
  title={Random forests},
  author={Breiman, Leo},
  journal={Machine learning},
  year={2001},
  publisher={Springer}
}

@inproceedings{grinsztajn2022why,
  title={Why do tree-based models still outperform deep learning on typical tabular data?},
  author={Grinsztajn, L{\'e}o and Oyallon, Edouard and Varoquaux, Ga{\"e}l},
  booktitle={Advances in Neural Information Processing Systems (NeurIPS)},
  year={2022}
}

@misc{brohan2023open,
      title={Open X-Embodiment: Robotic Learning Datasets and RT-X Models}, 
      author={Embodiment Collaboration and Abby O'Neill and Abdul Rehman and others},
      year={2023},
      eprint={2310.08864},
      archivePrefix={arXiv},
      primaryClass={cs.RO}
}

@misc{dulac2019challenges,
      title={Challenges of Real-World Reinforcement Learning}, 
      author={Gabriel Dulac-Arnold and Daniel Mankowitz and Todd Hester},
      year={2019},
      eprint={1904.12901},
      archivePrefix={arXiv},
      primaryClass={cs.LG}
}

@book{hastie2009elements,
  title={The Elements of Statistical Learning: Data Mining, Inference, and Prediction},
  author={Hastie, Trevor and Tibshirani, Robert and Friedman, Jerome},
  year={2009},
  publisher={Springer Science \& Business Media},
  edition={2nd}
}

@article{guyon2003introduction,
  title={An introduction to variable and feature selection},
  author={Guyon, Isabelle and Elisseeff, Andr{\'e}},
  journal={Journal of machine learning research},
  year={2003}
}

@inproceedings{louppe2013understanding,
  title={Understanding variable importances in forests of randomized trees},
  author={Louppe, Gilles and Wehenkel, Louis and Sutera, Antonio and Geurts, Pierre},
  booktitle={Proceedings of the 27th International Conference on Neural Information Processing Systems},
  year={2013}
}

@INPROCEEDINGS{tu2025rsizing,
  author={Tu, Jindong and Li, Yapeng and Xu, Peng and Li, Tuo and Li, Guoqing and Xie, Zushuai and Yu, Bei and Chen, Tinghuan},
  booktitle={2025 IEEE/ACM International Conference On Computer Aided Design (ICCAD)}, 
  title={RSizing: Robust Bayesian Optimization for Analog Circuit Sizing Under Process Variations}, 
  year={2025}
}

@inproceedings{ma2019cad,
  title={{CAD} Tool Design Space Exploration via Bayesian Optimization},
  author={Ma, Yuzhe and Yu, Ziyang and Yu, Bei},
  booktitle={ACM/IEEE Workshop on Machine Learning for CAD (MLCAD)},
  year={2019}
}

@inproceedings{eriksson2019scalable,
author = {Eriksson, David and Pearce, Michael and Gardner, Jacob R and Turner, Ryan and Poloczek, Matthias},
title = {Scalable global optimization via local Bayesian optimization},
year = {2019},
booktitle = {Proceedings of the 33rd International Conference on Neural Information Processing Systems}
}

@inproceedings{FLEX,
author = {Liu, Xingyu and Liang, Jiawei and Du, Linfeng and Zhang, Yipu and Ma, Chaofang and Fan, Hanwei and Xu, Jiang and Zhang, Wei},
booktitle = {Proceedings of the 54th International Conference on Parallel Processing (ICPP)},
title = {FLEX: Leveraging FPGA-CPU Synergy for Mixed-Cell-Height Legalization Acceleration},
year = {2025}
}

@misc{FPGN,
      title={FPGN: Redefining Ultra-Fast Programmable Gate-based Neural Acceleration with Differentiable LUTs}, 
      author={Jiawei Liang and Haotong Qin and Linfeng Du and Xingyu Liu and Shangkun Li and Hui Yu and Michele Magno and Xinyu Chen and Jiang Xu and Wei Zhang},
      year={2026},
      eprint={2607.08427},
      archivePrefix={arXiv},
      primaryClass={cs.AR},
      url={https://arxiv.org/abs/2607.08427}, 
}

@misc{zhang2026versaq3darchitecturesupportvisual,
      title={VersaQ-3D: Architecture Support for Visual Geometry Grounded Transformers via Versatile Quantization}, 
      author={Yipu Zhang and Jintao Cheng and Xingyu Liu and Zeyu Li and Carol Jingyi Li and Jin Wu and Lin Jiang and Ceyu Xu and Yuan Xie and Jiang Xu and Wei Zhang},
      year={2026},
      eprint={2601.20317},
      archivePrefix={arXiv},
      primaryClass={cs.AR},
      url={https://arxiv.org/abs/2601.20317}, 
}
\end{document}